\documentclass[aps,amsmath,twocolumn,pre]{revtex4}
\usepackage{amsmath}
\usepackage{amssymb}
\usepackage{epsfig}
\usepackage{amscd}
\usepackage{graphicx,xspace}
\usepackage{float}
\usepackage[normalem]{ulem}
\newcommand{\pmax}{p^{\max}}
\newcommand{\pitilde}{\tilde{\pi}}
\newcommand{\eqntext}[1]{ \left\{\!\!\mbox{\begin{tabular}{c}  #1 \end{tabular}}\!\! \right\}}

\newcommand{\flabel}[1]{\label{f:#1}}
\newcommand{\elabel}[1]{\label{e:#1}}
\newcommand{\slabel}[1]{\label{s:#1}}
\newcommand{\eq}[1]{eqn~(\ref{e:#1})}

\newcommand{\fig}[1]{Fig.~\ref{f:#1}}
 
\newcommand{\sect}[1]{Section~\ref{s:#1}} 
 
\newcommand{\ZZ}{\mathbb{Z}}  %
\newcommand{\SET}[1]{\{#1\}}
\newcommand{\expa}[1]{\mathrm{e}^{#1}}   
\newcommand{\expb}[1]{\exp \glb #1 \grb} 
\newcommand{\expc}[1]{\exp \glc #1 \grc} 
\newcommand{\expd}[1]{\exp \gld #1 \grd} 
\newcommand{\loga}[2][]{\log^{#1}\! \gla #2 \gra}  

\newcommand{\gla}{\,}  
\newcommand{\gra}{}  
\newcommand{\glb}{\left(}  
\newcommand{\grb}{\right)}  
\newcommand{\glc}{\left[}  
\newcommand{\grc}{\right]}  
\newcommand{\gld}{\left\{}  
\newcommand{\grd}{\right\}}  

\newcommand{\const}{\text{const}}
\newcommand{\TO}{,\ldots,}
\newcommand{\dd}[1]{\text{d}{#1\ }}   
\newcommand{\ddd}[1]{\text{d}{#1}}   

\newcommand{\mean}[1]{\left\langle #1 \right\rangle}
\newcommand{\half}{\frac{1}{2}}
\newcommand{\kB}{k_{\text{B}}}  

\newcommand{\rhomat}[4][]{\rho^{\text{#1}\!}\glb #2,#3,#4\grb}
\newcommand{\Rcut}{R^{\text{cut}}}  
\newcommand{\ksmall}{\ensuremath{k \ll \frac{L^2}{2\beta}}}  
\newcommand{\kbig}{\ensuremath{k \gg \frac{L^2}{2\beta}}} 
\newcommand{\Nexc}{N_{\text{exc}}}  
\newcommand{\Nsat}{N_{\text{sat}}}  

\begin{document}
\title{Off-diagonal long-range order, cycle probabilities, and condensate fraction\\
in the ideal Bose gas}
\author{Maguelonne Chevallier} \email{maguelonne.chevallier@ens.fr}
\author{Werner Krauth} \email{werner.krauth@ens.fr}
\affiliation{CNRS-Laboratoire de Physique Statistique \\ 
Ecole Normale
Sup{\'{e}}rieure; 24 rue Lhomond, 75231 Paris Cedex 05, France
}
\date{\today}

\begin{abstract}
We discuss the relationship between the cycle probabilities in the
path-integral representation of the ideal Bose gas, off-diagonal
long-range order, and Bose--Einstein condensation. Starting from
the Landsberg recursion relation for the canonic partition function,
we use elementary considerations to show that in a box of size $L^3$
the sum of the cycle probabilities of length $k \gg L^2 $ equals the
off-diagonal long-range order parameter in the thermodynamic limit.
For arbitrary systems of ideal bosons, the integer derivative of the
cycle probabilities is related to the probability of condensing $k$
bosons. We use this relation to derive the precise form of the $\pi_k$
in the thermodynamic limit.  We also determine the function $\pi_k$
for arbitrary systems.  Furthermore we use the cycle probabilities to
compute the probability distribution of the maximum-length cycles
both at $T=0$, where the ideal Bose gas reduces to the
study of random permutations, and at finite temperature.  We close
with comments on the cycle probabilities in interacting Bose gases.
\end{abstract}
\maketitle

\section{Introduction}
\slabel{introduction}
The canonic partition function of $N$ ideal bosons in a system
with energy levels $\epsilon_0 =0 < \epsilon_1 <
\dots$ at inverse temperature $\beta=1/(\kB T)$ is given by
\begin{equation}
   Z_N=\left[\prod_{\sigma}\sum_{n_{\sigma}=0}^{N}\right]
       \expb{-\beta\sum_{\sigma}n_{\sigma}
       \epsilon_{\sigma}} \delta_{\sum_{\sigma}n_{\sigma},N}.
   \elabel{Z_def_general}
\end{equation}
For each state $\sigma$, the allowed occupation numbers $n_{\sigma}$
go from $0$ to $N$, and the Kronecker $\delta$-function enforces the
total number of particles to be $N$.

The Feynman path-integral \cite{feynman} represents $Z_N$ as a trace
over the diagonal density matrix $\rho^{\text{bos}}$,
\begin{equation*}
   Z_N = \int  \ddd{x_1} \dots \ddd{x_N}  
   \rhomat[bos]{\SET{x_1 \TO x_N}}{\SET{x_1 \TO x_N}}{\beta}, 
\end{equation*} 
which is  given in terms of the $N!$ permutations $P$ of the
distinguishable-particle density matrix $\rho$
\begin{multline}
\rhomat[bos]{\SET{x_1 \TO x_N}}{\SET{x'_1 \TO x'_N}}{\beta} \\
= \frac{1}{N!}
\sum_P \rhomat{\SET{x_1 \TO x_N}}{\SET{x'_{P_1} \TO x_{P_N}'}}{\beta}.
\elabel{bos_dist_dens_mat}
\end{multline}
Thus the partition function is a sum of $N!$ permutation-dependent terms, 
$Z_N = \sum_P Z_P$, and it is possible to define the probability of 
a permutation $P$ as
\begin{equation}
\pi_P = \frac{Z_P}{Z_N}.
\elabel{Z_permutation}
\end{equation}
Permutations can be broken up into cycles, and one may also define cycle probabilities:
\begin{equation}
\pi_{k} = \frac{1}{Z_N}\sum_{P \in \SET{N,k}}Z_{P} , 
\elabel{cycle_probabilities}
\end{equation}
where $\SET{N,k}$ denotes the set of all permutations where the particle
$N$ belongs to a cycle of length $k$.  The choice of the
particle $N$ in \eq{cycle_probabilities} is arbitrary.

In this paper,
we are concerned with the characterization of the cycle probabilities
$\pi_k$ in the ideal Bose gas and with their relation to quantities
characterizing Bose--Einstein condensation, namely the off-diagonal long 
range order
(in \sect{odlro}, see also \cite{ueltschi,benfatto}) and the condensate
fraction (in \sect{derivative}). In fact, for arbitrary finite systems of
ideal bosons, the discrete derivative of the function $\pi_k$ yields the
probabilities for condensing $k$ bosons \cite{holzmann}.  This relation
between the cycle statistics and the condensate fraction has not been
scrutinized before in detail. It allows to determine the $\pi_k$ from
the known fluctuation properties of the ideal Bose gas.  It is also
possible, and very instructive, to compute the cycle probabilities $\pi_k$
directly via the infinite-density limit of a finite Bose gas
(see \sect{p_k}).  Furthermore, we will exploit some of the subtleties of
the concept of cycle probabilities to compute the probability
distribution of the maximum-length cycle at zero temperature, where
the problem reduces to a study of random permutations \cite{shepp}.
Our preceding analysis of the $\pi_k$ allows us to understand why this
distribution is essentially unchanged at finite temperature, below
the condensation temperature (\sect{random_permutations}).  Finally,
we briefly review some known relations between Bose condensation
and the presence of infinite cycles for interacting Bose gases
(\sect{interacting}).

\section{Recursion relations for the partition function and the density matrix}
\slabel{landsberg}
The Landsberg recursion relation \cite{landsberg} gives $Z_N$ as
a sum of only $N$ terms 
\begin{equation}
   Z_N=\frac{1}{N}\sum_{k=1}^N Z_{N-k}z_k,
   \elabel{partition_function_rec}
\end{equation}
where the $z_k = Z_1(k\beta)= \sum_{\sigma} \expb{-k \beta
\epsilon_{\sigma}}$ are the single-particle partition functions at inverse
temperature $k \beta$.  In fact, the terms appearing in \eq{partition_function_rec} 
are the cycle probabilities:
\begin{equation}
\pi_{k} =Z_{N-k}z_{k}/(N Z_N).
\elabel{cycle_prob_definition}
\end{equation}
This is because $k$ particles  on a permutation cycle of length $k$
at inverse temperature $\beta$ have the same statistical weight $z_k$ as
a single particle at inverse temperature $k \beta$ and, furthermore, in the
ideal Bose gas, the different cycles of a permutation are statistically
independent. Thus, in \eq{cycle_prob_definition} the
cycle contributes $z_k$ and the remaining $N-k$ particles contribute
$Z_{N-k}$ (see \cite{SMAC}).

The off-diagonal single-particle density matrix
\begin{multline*}
   \rho_N(r,r',\beta) = \int  \ddd{x_1} \dots \ddd{x_{N-1}}   \\
\times \rhomat[bos]{\SET{x_1 \TO x_{N-1},r}}{\SET{x_1 \TO x_{N-1},r'}}{\beta},
\end{multline*}   
is also a sum of permutation-dependent terms, and the Landsberg recursion
relation can be generalized to a non-diagonal single-particle density
matrix $\rho(r,r',k\beta)$ rather than to the diagonal one:
\begin{align}
  \rho_N(r,r',\beta) &= \frac{1}{N}\sum_{k=1}^{N} Z_{N-k} \rho(r,r',k\beta), 
    \elabel{rho_sum}\\
\frac{V}{Z_N}\rho_N(r,r',\beta)   &= \sum_{k=1}^{N} \pi_k \Rcut_k(r,r'),
    \elabel{off_diag_def} 
\end{align}
where $V$ is the volume of the system. 
In \eq{off_diag_def}, the cycle probabilities $\pi_k$ are
modified by the cut-off function $\Rcut_k(r,r')$, which is
proportional to the ratio of the off-diagonal and the diagonal
density matrices.  In the three-dimensional homogeneous
Bose gas in a periodic cubic box of size $L^3$, we have
$\Rcut_k(r,r')=L^3\rho(r,r',k\beta)/z_k=\rho(0,r-r',k\beta)/\rho(0,0,k\beta)$.

As is well known, the condensate fraction is proportional to
the off-diagonal single-particle density matrix in the limit
$|r-r'|\rightarrow\infty$, a case that corresponds to $|r-r'|\propto L$
in a cubic box of length $L$ for $L \rightarrow \infty$.  The cut-off
function then  vanishes for small $k$, and terms with $\kbig$\ dominate
the sum in \eq{off_diag_def} \cite{footnote_notation}.

\begin{figure}
   \centerline{
   \epsfxsize=9.0cm
   \epsfbox{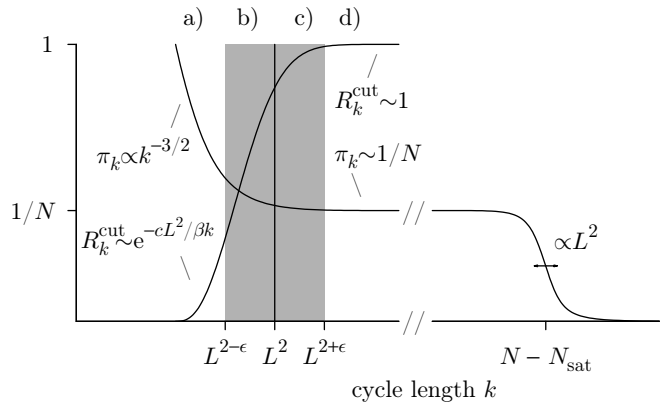}}
   \caption{Cycle probabilities $\pi_k$ and cut-off function $\Rcut_k$
   for a periodic box of size $L^3$ below $T_c$.}
   \flabel{cut_off}
\end{figure}

\section{Off-diagonal long-range order and cycle probabilities}
\slabel{odlro}

In this section, we discuss a relation between the off-diagonal
long-range order parameter and the sum of cycle probabilities for the
homogeneous Bose gas, with unit mass and  particle density ($m=\hbar=1$,
$N=L^3, T_c=2\pi/\zeta(3/2)^{2/3}$).
To analyze \eq{off_diag_def}, we notice that $z_{k}\sim 1$ for \kbig\
and $z_{k}\sim L^{3}/(2\pi\beta k)^{3/2}$ for \ksmall.  It follows from
$Z_{N-k} < Z_N$ that
\begin{equation}
\pi_k \lesssim
\begin{cases}
1/(2\pi\beta k)^{3/2}\ &\text{for}\  \ksmall \\
1/N\ &\text{for}\  \kbig.
\end{cases}
\elabel{simple_pi_k_conditions}
\end{equation}

We now study the cut-off function $\Rcut_{k}(r,r')$ for $|r-r'|
\propto L$.  For concreteness, we suppose that the vector $r-r'$ has
the same components in all three space directions: $r-r'= L(\delta,
\delta, \delta)$ with $0<\delta< \half$ and denote the corresponding cut-off
function and the off-diagonal density matrix by $\Rcut_{k}(\delta)$
and by $\rho_N(\delta)$, respectively. We have
\begin{equation*}
   \Rcut_{k}(\delta)=\glc \frac{\sum_{w\in \ZZ}e^{-(\delta+w)^2 L^{2}/(2k\beta)}}
   {\sum_{w\in \ZZ}e^{-w^2 L^{2}/(2k\beta)}} \grc^3 .
\end{equation*}
For \kbig, both the numerator and the  denominator are $\sim
\sqrt{2\pi\beta k}/L$, so that $\Rcut_{k}(\delta)\sim 1$. For \ksmall,
the numerator is $\sim\expd{-3\delta^{2}L^{2}/(2\beta k)}$, and
the denominator is $\sim 1$.  This shows that the monotonic function
$\Rcut_k(\delta)$ is exponentially small for \ksmall\ and equal to unity
for \kbig\ (see \fig{cut_off}).  The estimates for $\Rcut_k$ and for $\pi_k$ imply
\begin{equation}
\begin{split}
&\lim_{L \rightarrow \infty}
\sum_{k=1}^{L^2}\pi_{k}\Rcut_{k}(\delta)=0, \\
&\lim_{L \rightarrow \infty}\sum_{k= L^2}^{N}\pi_{k}
\glc 1-\Rcut_{k}(\delta) \grc = 0.
\end{split}
\elabel{product_cut_off_pi}
\end{equation}
(see the appendix for a short, but rigorous derivation).
It follows that
\begin{equation}
   \lim_{L\rightarrow\infty}
   \glc \frac{L^{3}\rho_N(\delta)}{Z_{N}}-
   \sum_{k= \const L^{2}}^{N}\pi_{k} \grc=0.
   \elabel{egalite}
\end{equation}
This equation relates the probabilities of infinite
cycles to off-diagonal long-range order and (because
$\lim_{|r-r'|\rightarrow\infty}\frac{V}{Z_N}\rho_N(r,r',\beta)$ gives the
condensate fraction) also to the condensate fraction.  The relation 
between cycle lengths and off-diagonal long-range order is natural
because, in order for a cycle to contribute to the off-diagonal density
matrix for $| r - r' | \propto L$, its de Broglie wavelength must be at
least comparable to the distance $L$, which means that the particle must
belong to a cycle of length $k$ with $L\lesssim\sqrt{2 \pi k \beta}$
(see also \cite{suto,ueltschi} for related derivations).

\section{Condensate fraction as a derivative of the cycle probabilities}
\slabel{derivative}

In \sect{odlro}, we studied the sum of the cycle probabilities. We now
discuss the fact that the discrete derivative of the cycle probabilities
exactly gives the condensation probabilities.  This relation allows to
compute the $\pi_k$, and it involves neither a cut-off function nor the
thermodynamic limit and holds for arbitrary systems of ideal bosons.
The relation was mentioned very briefly in the context of Monte Carlo
calculations \cite{holzmann}, but it was not analyzed in any detail.

Let us define $S_{N}(k,\sigma)$ as the canonic partition function with $N$
bosons of which exactly $k$ are in state $\sigma$ (see \cite{landsberg})
and, similarly, the restricted partition function with at least
$k$ bosons in the state $\sigma$,  as
\begin{equation}
   Y_N(k,\sigma)=\sum_{k' = k }^N S_N(k',\sigma) = 
   \expa{-\beta k \epsilon_{\sigma}} Z_{N-k}.
   \elabel{SetZ}
\end{equation}
($S_{N}(k,\sigma)/Z_N$ is the probability of having exactly $k$ bosons
in state $\sigma$.)  Summing \eq{SetZ} over all states $\sigma$, and
dividing by $N Z_N$, we recover the cycle probabilities,
\begin{equation*}
   \frac{1}{N Z_N}\sum_\sigma Y_N(k,\sigma) =
   \frac{\sum_{\sigma}\expa{-\beta k 
   \epsilon_{\sigma}}}{N Z_N} Z_{N-k}= \pi_k, 
\end{equation*}
whose negative discrete derivative, 
\begin{equation}
   \pi_{k}-\pi_{k+1}=\sum_\sigma S_N(k,\sigma)/(N Z_N),
   \elabel{discrete_derivative}
\end{equation}
thus yields the sum over all $\sigma$ of the probabilities of
condensing $k $ bosons into state $\sigma$. We note that the l.h.s. of
\eq{discrete_derivative} contains only quantities related to the
path-integral picture, whereas the r.h.s. involves only energy levels.
Furthermore, we note that the probability of condensing $k$ particles into
excited levels is zero for large $k$, so that \eq{discrete_derivative}
effectively relates the groundstate condensation probabilities to the
derivative of the cycle probabilities.
\begin{figure}
   \centerline{
   \epsfbox{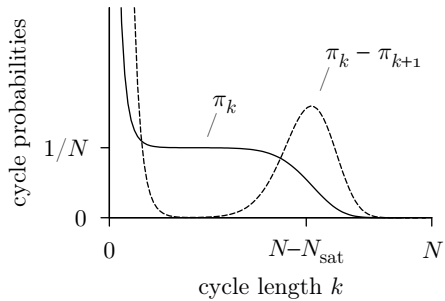}}
   \caption{Cycle probabilities $\pi_k$ and their negative discrete
   derivative (not to scale) for $N=125$ ideal bosons in a periodic box
   of size $L^3=5^3$ at temperature $T/T_c=0.5$ ($\Nsat=48.7$).}
   \flabel{pi_k_deriv}
\end{figure}

To illustrate the relation between cycles and condensation probabilities,
we consider the example of \fig{pi_k_deriv} of $N$ particles in a
periodic box, with the cycle probabilities explicitly computed from the
Landsberg recursion relation.  For $k>N/2$, the derivative of the cycle
probabilities equals the sum of the probabilities of condensing $k$
bosons into state $\sigma$. More generally, for \kbig, the condensation
probabilities into the excited states vanish and  $\pi_{k}-\pi_{k+1}\sim
S_N(k,0)/(N Z_N)$ directly yields the probability distribution of having
$k$ particles in the groundstate, a distribution with mean value $k =
\mean{N_0}$ and standard deviation $\propto L^2$.  On the other hand,
for small $k$, as already discussed, the discrete derivative behaves
as $k^{-5/2}$. It describes the probabilities of condensing $k$ bosons
into excited states.  Finally, for $ \frac{L^2}{2\beta} \ll k$ and $k
\le \mean{N_0}-L^2/(2\pi\beta)^{3/2}$, the probability of having $k$
particles in the condensate vanishes, so that the $\pi_k$ are constant.

\section{Direct calculation of cycle probabilities}
\slabel{p_k}

We have seen in \sect{derivative} that the cycle probabilities $\pi_k$ follow
from the known fluctuations of the condensate in the canonic ensemble.
It is very instructive to compute them directly via the infinite-density
limit for the partition function of an arbitrary system.
For $\kbig$, we have $z_k \sim 1$, so that $\pi_k \sim  Z_{N-k}/Z_N$.
For illustration, the \fig{Z_N} shows the $Z_k$  
in a fixed physical system  (for the homogeneous Bose
gas, in a fixed box of size $L^3$), as it can be computed 
numerically from the Landsberg recursion relation.  
The limiting value $Z_\infty$ (at infinite density) and 
the value $k=\Nsat$ of  largest variation of $Z_k$ can be computed exactly.

Indeed, as the groundstate energy is zero, $ Z_{N}$ differs from $Z_{N-1}$ only by
configurations without any particle in the groundstate: $Z_N-Z_{N-1}=
S_{N}(0,0)$ (see above \eq{SetZ}). Therefore, $Z_{N}=\sum_{k=1}^{N}S_{k}(0,0)$ with
\begin{equation*}
   S_{k}(0,0)=\left[\prod_{\sigma > 0}\sum_{n_{\sigma}=0}^{\infty}\right]
   \expb{-\beta\sum_{\sigma > 0 }n_{\sigma}\epsilon_{\sigma}}
   \delta_{\sum_{\sigma > 0 }n_{\sigma},k}.
\end{equation*}
Summing this equation over all integers $k$ yields,  
for arbitrary states $\SET{\epsilon_{\sigma}}$,
\begin{equation}
   Z_{\infty}=\prod_{\sigma > 0}\frac{1}{1-\expb{-\beta\epsilon_{\sigma}}}=
   \expb{\sum_{k=1}^{\infty}\frac{z_{k}-1}{k}}.
   \elabel{Z_infty}
\end{equation}
This exact formula for a finite canonical Bose system can be
obtained without the usual saddle-point integration.  It agrees with
the grand-canonical partition function for the excited states at zero
chemical potential because at high density the condensate serves as a
reservoir for the excited bosons.

To determine the value of $N$ for which the partition function $Z_k$ 
passes from $Z_k \sim 0$ to $Z_k \sim Z_{\infty}$ (see \fig{Z_N}), we again 
take a discrete derivative and consider the probability distribution
of the number of  excited particles
\begin{equation*}
   \pi_{\infty}(\Nexc=k)=(Z_{k}-Z_{k-1})/Z_{\infty},
\end{equation*}
whose mean value, the saturation number, is 
\begin{multline}
\elabel{Nsat}
\Nsat=\mean{\Nexc}=\sum_{k=1}^{\infty}k \pi_{\infty}(\Nexc=k)=\\
\sum_{\sigma > 0 }
\frac{\expb{-\beta\epsilon_\sigma}}
{1-\expb{-\beta\epsilon_\sigma}} = \sum_{k=1}^{\infty}(z_{k}-1).
\end{multline}
The variance of the number of excited particles is again given exactly by 
the corresponding grand-canonical expressions
\begin{multline}
\mean{\Nexc^2} - \mean{\Nexc}^2=\sum_{\sigma > 0}\frac{1}{\expb{\beta\epsilon_{\sigma}}-1}+\frac{1}   {\glc\expb{\beta\epsilon_{\sigma}}-1 \grc^{2}}\\
=\Nsat+\sum_{k=2}^{\infty}(k-1)(z_{k}-1).
\elabel{varNsat}
\end{multline}

\begin{figure}
   \centerline{
   \epsfxsize=4.5cm
   \epsfbox{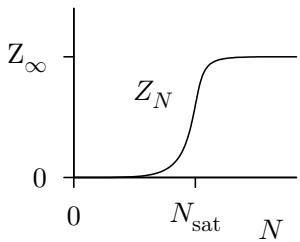}}
   \caption{Canonic partition function $Z_N$ as a function of $N$ in a finite system. }
   \flabel{Z_N}
\end{figure}
For completeness, we give the expressions corresponding
to \eq{Z_infty}--\eq{varNsat} in the homogeneous Bose gas, in the limit
$L\rightarrow\infty$: $Z_\infty \sim \expc{ L^3/(2\pi
\beta)^{-5/2}\zeta(5/2) }$ and, furthermore, $\Nsat/L^3\sim
\zeta(3/2)/(2\pi\beta)^{3/2}=\eta_c$, with $\eta_c$ the critical density.
In addition, the variance of $\Nexc$ behaves as $L^4/(2\pi\beta)^3$
\cite{ziff}, so that the jump of $Z_k/Z_N$ from $0$ to $1$, for $T <
T_c$, takes place in a window of width $L^2$, as already noted in
\sect{derivative}.

Putting together the analysis of $Z_k$ for a fixed size of the system
and a varying particle number, and our knowledge of $z_k$, we have a
complete description of the cycle probabilities $\pi_k$.
Once we know exactly the behavior of the $\pi_k$
in the thermodynamic limit, we observe that the sum of the $\pi_k$ for
$\ksmall$ gives exactly the critical density $\eta_c$, and the sum of
the $\pi_k$ for $\kbig$ gives the condensate fraction. This conclusion
is the same as in \sect{odlro}, but it stems from a microscopic analysis
of the cycle probabilities.

\section{Maximum cycle lengths}
\slabel{random_permutations}

At finite temperature, the cycle probabilities $\pi_k$ are easily
computed from the Landsberg recursion relation, and at $T=0$, they
are given by $\pi_k=1/N$.  In Monte Carlo calculations the entire
permutation of $N$ elements can be sampled,  starting from the cycle
containing the element $N$, which has length $k$ with probability
$\pi_k$. This leaves one with a system of $N-k$ particles, for which
the probabilities of the cycle containing element $(N-k)$ can again be
computed, etc.  (see \cite{SMAC}). (At zero temperature, the distribution
remains $ \pi_k = 1/(N-k)$, at finite temperature, it is given by
\eq{cycle_prob_definition}.)

In this section, we use the cycle probabilities $\pi_k$ to obtain
analytical results for $\pmax_k$, the probability that the longest cycle
in a permutation has length $k$.  This probability distribution is closely
related to the distribution $\pi_k$ both at finite temperature and at
$T=0$, where the ideal Bose gas is equivalent to the problem of random
permutations, which has been much studied in the mathematics literature.
We recover some classic results which we  generalize to 
finite temperatures.

We first note that the cycle probabilities are related to the mean
number of cycles of length $k$, $\pitilde_k= N\pi_k/k$.  For $k > N/2$,
$\pitilde_k$ coincides with the probability $p_k$ that the permutation
$P$ has at least one cycle of length $k$. This is simply because a
permutation of $N$ elements can have no more than one such cycle, in
other words, because

\begin{gather}
\pitilde_k = \sum_{m=1}^\infty m \times 
\eqntext{prob. to have \\ $m$ cycles of length $k$}  
\elabel{mean_cycle_k}
\\
\intertext{and}
p_k = \sum_{m=1}^\infty  
\eqntext{prob. to have \\ $m$ cycles of length $k$} ,
\elabel{prob_cycle_k}
\end{gather}
and only terms with $m=1$ contribute to the above expressions if $k>N/2$.

Furthermore, any cycle of length $k> N/2$ must be the longest cycle, 
so that we arrive for all temperatures at:
\begin{equation}
\pmax_k = p_k = \pitilde_k = N \pi_k/k \quad \text{for}\ k > N/2, \forall T.
\elabel{explicit_all_T}
\end{equation}
We next consider the probability distributions $p_k$ and $\pmax_k$ for $k
\le N/2$ at $T=0$, for $N \rightarrow \infty$. We take the continuum
limit $p_k \to p(x)$ by setting $p_k=\frac{1}{N}p(x=\frac{k}{N})$ 
(and similarly for $\pmax(x)$ and $\pitilde(x)$).

It follows from the recursive procedure, and from the fact that the
$\pi_k$ are of order $1/N$, that the probability to have more than one
cycle of the same length is $O(1/N^2)$. In the limit $N \rightarrow
\infty$, we see that $p(x) = \pitilde(x) = 1/x$.

For $x<1/2$, the probability $\pmax(x)$ is the product of the probability
to sample $x$, and the probability that $x$ remains the longest cycle
in the remaining partition of length $1-x$, in other words:
\begin{equation*}
\pmax(x)= p(x) \glb 1 - \eqntext{prob. that longest cycle in \\
                                         $[0,1-x]$ is $ > x$} \grb .
\end{equation*}
At $T=0$, and in the limit $N \rightarrow \infty$, the probability
that in a permutation of length $N(1-x)$ the longest cycle's length exceeds $Nx$
equals the probability that in a permutation of $N$ elements the longest cycle 
exceeds  $x/(1-x)$, 
and we arrive at
\begin{equation}
   \pmax(x)= p(x) \glc 1 - \int_{x/(1-x)}^1 \dd{x'} \pmax(x')  \grc. 
   \elabel{f_max_iteration}
\end{equation}
For all $x\in]0,1[$, $x < x/(1-x)$, so \eq{f_max_iteration} can be used
to compute $\pmax(x)$ from $x=1$ downwards to arbitrary precision 
(see \fig{partition}).
Alternatively, we can transform \eq{f_max_iteration} into a differential equation
\begin{equation*}
\frac{d}{dx} \pmax(x) = - \frac{1}{x} \pmax(x) + \frac{1}{x(1-x)^2} 
\pmax\glb \frac{x}{1-x} \grb .
\elabel{differential_equation}
\end{equation*}
Explicit formulas for $\pmax(x)$ are easily obtained, starting by
inputting, from  \eq{explicit_all_T}, $\pmax(x)$ for $\half < x<1$
to obtain $\pmax$ in the window $x \in [\frac{1}{3}, \half]$, etc,
\begin{equation}
\pmax(x) =
\begin{cases}
\frac{1}{x}\ \text{for $\half < x $}\\
\frac{1}{x} \glc 1 - \log\glb \frac{1-x}{x} \grb \grc  
        \text{for $\frac{1}{3} < x < \half$} ,
\end{cases}
\elabel{}
\end{equation}
but they become cumbersome  (see also \cite{goncharov}).

At finite temperature, the cycle of the element $N$ is sampled from a
distribution $\pi_k$ as shown in \fig{pi_k_deriv} rather than from the
flat distribution. However, the longest cycle is again of length $\propto
N$, and it is sampled from an essentially constant distribution in the
interval $x \in [0, \mean{N_0}/N]$. In the thermodynamic limit, for $T <
T_c$, the distribution $\pmax(x)$ of the properly rescaled variable $x =
k^{\max}/\mean{N_0}$ trivially agrees with the one obtained at $T=0$.
In particular, the mean length of the longest cycle is given by $0.624
\mean{N_0}$ at all temperatures $T < T_c$ \cite{shepp}.  At sufficiently
low temperature, the rounding at $ x \simeq 1$ is \emph{exactly} described
by \eq{explicit_all_T} (see \fig{partition}).

\begin{figure}[htbp]
\begin{center}
\includegraphics{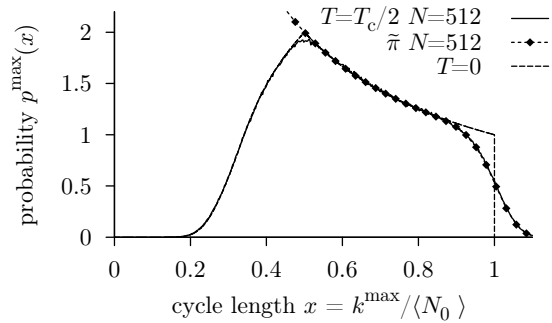}
\end{center}
\caption
{Probability $\pmax(x)$ from a numerical integration of
\eq{f_max_iteration} at $T=0$, compared to the distribution $\pmax_k$ obtained from
a Monte Carlo sampling at  $T=T_c/2$, and the exact expression of \eq{explicit_all_T}
for large $x$, at $T=T_c/2$.}
\flabel{partition}
\end{figure}
\section{Interacting Bose gas}
\slabel{interacting}

For the interacting Bose gas, the relation between off-diagonal
long-range order and the infinite cycles has been studied for generalized
mean-field models (\cite{dorlas,ueltschi}), but
it has proven difficult to establish rigorous results for models with
realistic interactions.  In the general case, the definition
of the superfluid density in terms of the winding numbers \cite{pollock87}
proves that for the homogeneous Bose gas in three dimensions, the presence
of infinite cycles is equivalent to a non-zero superfluid density.
It is generally admitted that in homogeneous systems, Bose--Einstein
condensation implies superfluidity, so that the presence of a
condensate would imply the existence of infinite cycles in those systems.


On the other hand, the direct link between the cycle probabilities and the
condensate fraction, as described in \sect{derivative}, cannot hold for
interacting systems. This is easily seen at $T=0$ because the groundstate
of any system of identical but distinguishable particles has bosonic
symmetry. Therefore, the distinguishable-particle density matrix 
\begin{multline*}
\rhomat{\SET{x_1 \TO x_N}}{\SET{x_{P_1} \TO x_{P_N}}}{\beta= \infty}
\end{multline*}
is independent of the permutation $P$. It follows that at $T=0$, and
for finite $N$, all permutations have the same weight and the cycle
probabilities again satisfy $\pi_k=\frac{1}{N}$. At the same time,
the condensate fraction of interacting systems at $T=0$ is less than
one\cite{Penrose}, and the distribution of the $\pi_k$ cannot be related
to the distribution of  $N_0$ in the same manner as in the ideal Bose gas.

\section{Conclusion}
\slabel{conclusion}
In conclusion, we have discussed the relation between the off-diagonal
long-range order, the condensate fraction, and the cycle probabilities in
the ideal Bose gas. Only long cycles ($k > \const L^2$  in the homogeneous
case) contribute to the off-diagonal long-range order parameter, and Bose
Einstein condensation is equivalent to the presence of infinite cycles.
We have also discussed the probability distribution for long cycles,
and the general, non-intuitive,  link between the integer derivative of
the cycle probabilities and the number of condensed bosons.  This integer
derivative provides us with a purely topological characterization of the
condensate fraction in the ideal Bose gas. Our knowledge of the cycle
probabilities allows us to study the distribution of the longest cycle. We
have remarked that the $0<T<T_c$ case can be understand from the $T=0$
case. The mean length of the longest cycle is proportional to the number
of condensed bosons. For the interacting Bose gas, at the present time,
only the winding-number formula\cite{pollock87} has been rigorously shown
to relate the topology of Feynman paths to another characteristic of
interacting Bose-condensed systems, namely the superfluid density. The
link between Bose--Einstein condensation and superfluidity will need
to be explored further in order to better understand the cycles of the
interacting Bose gas.

\acknowledgments
We thank A. Comtet and S. N. Majumdar for an inspiring discussion about
maximum cycle length (see also \cite{comtet_majumdar}) and for making
us aware of the papers \cite{shepp,goncharov}.

\appendix
\slabel{appendix}
\section{Derivation of \eq{product_cut_off_pi}}
\slabel{derivation}

In this appendix, we prove \eq{product_cut_off_pi}. We consider the sums
separately in the regions a) -- d) indicated in \fig{cut_off} and show
that they all vanish individually.

In region a), $\pi_k\Rcut_{k}(\delta)\lesssim
\pi_{k}\expc{-3\delta^{2}L^\epsilon/\beta}$, and the normalization of
the $\pi_{k}$ shows that the partial sum vanishes
in the limit $L\rightarrow\infty$. In b), $\pi_{k}\Rcut_{k}\leq
\pi_{L^{2-\epsilon}}\Rcut_{L^2}\lesssim \const/L^{3(2-\epsilon)/2}$ and
the partial sum is smaller than $\const L^{3\epsilon/2-1}$. In c), as
$1-\Rcut_{k}(\delta)$ decreases with $k$, $\pi_{k}\glc 1-\Rcut_{k}(\delta)
\grc \leq \pi_{L^{2}} \glc 1-\Rcut_{L^{2}}(\delta) \grc$ and the partial
sum is dominated by $\const L^{\epsilon-1}$. Finally, in d), $\pi_{k}\glc
1-\Rcut_{k}(\delta)\grc \leq \pi_{k}\glc 1-\Rcut_{L^{2+\epsilon}}(\delta)
\grc$. The normalization condition of the $\pi_{k}$ and the relation
$L^{2+\epsilon} \gg L^{2}$ imply that this partial sum vanishes for
$L\rightarrow\infty$.  With a suitable choice of $\epsilon$ (for example
$\epsilon = \half$), these sums all vanish for $L\rightarrow\infty$.
In fact, any function $\epsilon(L) \gg 1/\loga{L} $ that satisfies
$\epsilon(L)<2/3$ for $L>L_{0}$ can be used.

  
\end{document}